\documentclass[onecolumn,prb,superscriptaddress,longbibliography]{revtex4-2}
\usepackage{graphicx}
\usepackage{color}
\usepackage{amsmath}
\usepackage{enumitem}
\usepackage{amssymb}
\usepackage{hyperref}
\usepackage[normalem]{ulem}
\usepackage{adjustbox}

\hypersetup{
	colorlinks = true,
	linkcolor = blue,
	citecolor = blue,
	urlcolor  = blue,
}
\newcommand{\be}{\begin{equation}}
\newcommand{\ee}{\end{equation}}

\newcommand{\bea}{\begin{eqnarray}}
\newcommand{\eea}{\end{eqnarray}}

\newcommand{\p}{\partial}

\renewcommand{\Re}{{\rm \, Re\,}}

\renewcommand{\vec}[1]{{\boldsymbol #1}}
\renewcommand{\epsilon}{\varepsilon}

\def\nn{\nonumber\\}

\begin{document}
\title{Differential entropy per particle as a probe of van Hove singularities and flat bands}
\date{\today}

\author{Yelizaveta Kulynych}
\affiliation{Department of Physics, Taras Shevchenko National University of Kyiv, Kyiv 03680, Ukraine}

\author{D. O. Oriekhov}
\affiliation{Instituut-Lorentz, Universiteit Leiden, P.O. Box 9506, 2300 RA Leiden, The Netherlands}

\begin{abstract}
 In the present paper we derive the general expressions for the differential entropy per particle near van Hove singularities (vHs) in the density of states. The dependence of entropy per particle on chemical potential and temperature demonstrates different behavior depending on the type of vHs, and distinguishes high-order vHs with different divergence exponents and flat bands. In addition, it allows one to test the ``flatness" of the band in experiment. We compare the analytic predictions with the numerical calculation of the differential entropy for tight-binding models of graphene, Lieb lattice and square-octagon lattice. Our results show that the obtained analytic expressions capture the main features of the differential entropy, thus serving as a good probe for details of the density of states structure.  
 \end{abstract}
\maketitle

\section{Introduction}
Entropy is a fundamental thermodynamic quantity that describes the state of many-body system and governs its' thermoelectric, thermomagnetic and heat transport properties. However, the total entropy of the system is hard to be measured directly, while the differential entropy per particle is accessible for modern experimental techniques \cite{Kuntsevich2015}. The advantage of experiments preformed in Ref.\cite{Kuntsevich2015} is the design of the form of flat capacitor with proper modulation of temperature, which is particularly suitable for studying properties of 2D materials. The differential entropy is calculated through Maxwell relation $s=\left(\partial S/\partial n\right)_{T}=-\left(\partial \mu/\partial T\right)_{n}$ using the experimental data for evaluation of second derivative.  Recently, another set of experiments measuring differential entropy attracted great attention due to the observation of isospin Pomeranchuk effect in twisted bilayer graphene \cite{Saito2021Nature,Rozen2021Nature}. In this set of experiments the differential entropy was extracted from the transport data.

The differential entropy per particle was proven to be a useful tool to probe the peculiar features of band structure \cite{Tsaran2017Nature} and to show the presence of Lifshitz transitions with changing Fermi level \cite{Galperin2018}. In Ref.\cite{Varlamov2016} it was shown that for two-dimensional electronic gas with parabolic dispersion the entropy per electron, s, exhibits quantized peaks which correspond to resonance positions of chemical potential at size quantization levels. The universal structure of these peaks was linked \cite{Varlamov2016} to the topological Lifshitz transitions \cite{Lifshitz1960}. Later is was shown \cite{Tsaran2017Nature} that the entropy spikes in Dirac materials also correspond to the Lifshitz transitions, which occur at the boundaries of gapped bands. 

The above mentioned peaks and spikes of differential entropy \cite{Galperin2018,Varlamov2016,Tsaran2017Nature} are the clear signatures of regions with vanishing density of states in the band structure. The goal of the present paper is to analyze the character of the differential entropy per particle in the opposite case - near the regions in band structure with diverging density of states (DoS). The classic examples of such divergence of DoS are the van Hove singularities (vHs) \cite{vanHove1953} with logarithmic divergence and flat bands \cite{Sutherland1986,Shen2010PRB} with delta-function-like divergence \cite{Leykam2018}. Recently, a new class of DoS singularities was introduced - high-order van Hove singularities \cite{Yuan2019Nature}, which corresponds to power-law divergence of DoS. The corresponding band structure contains the so-called high-order saddle point at the energy level of this vHs \cite{Yuan2020PRB,Chandrasekaran2020PRB}. While usual saddle points occur in most two-dimensional materials due to periodicity of the Brillouin zone \cite{vanHove1953}, the high-order saddle points became a common feature of novel two-dimensional materials \cite{Yuan2019Nature}. Several examples of  such materials that host high-order vHs of different kinds are: $\beta-\text{YbAlB}_{4}$ \cite{Ramires2012PRL}, bilayer graphene with tuned dispersion via adding interlayer voltage bias \cite{Shtyk2017} and $\mathrm{Sr}_{3} \mathrm{Ru}_{2} \mathrm{O}_{7}$ \cite{Efremov2019}.
	Recently it was also shown, that the tight-binding model of monolayer graphene can host high-order vHs if the next- and next-next-nearest neighbor parameters are included and tuned to special critical values \cite{Classen2020}. 
	Notably, the novel experiment has shown that the energy level of the usual van Hove singularity in monolayer graphene can be accessed using the special technique of very high doping \cite{Rosenzweig2020PRL}. 
	
	The appearance of van Hove singularities near the Fermi level can lead to different prominent phenomena. 
	One expected phenomenon is the chiral superconductivity in monolayer graphene, that is predicted to occur at the level of vHs \cite{Nandkishore_2012Nature}. When a high-order vHs is placed close to the Fermi level, one can expect the enhancement of density
	wave and Pomeranchuk orders together with the superconductivity 
	\cite{Classen2020}. The role of high-order vHs on different pairing types in twisted bilayer graphene was analyzed in Ref.\cite{Lin2020arxiv}. For example, their presence in twisted bilayer graphene \cite{Sherkunov2018} can lead to valley magnetism \cite{Chichinadze2020}, density waves and unconventional superconductivity \cite{Isobe2018PRX} such as topological superconductivity \cite{Wang2020arxiv}, the ``high-T$_{c}$" phase diagram \cite{Lin2019highTc}, and Kohn-Luttinger superconductivity \cite{Gonzalez2019}.

In the recent papers \cite{Yuan2020PRB} and \cite{Chandrasekaran2020PRB} the classification of possible high-order van Hove singularities in 2D lattices with different symmetry groups was performed. These results motivate us to analyze the differential entropy as an experimentally-accessible quantity that can distinguish between different types of van Hove singularities. In the main text we derive the exact analytic expressions for the entropy per particle near usual logarithmic and high-order van Hove singularities, and completely flat bands using the effective models for density of states \cite{Yuan2019Nature,Yuan2020PRB,Chandrasekaran2020PRB}. Our expressions cover the whole range of known high-order van Hove singularities in 2D materials from Refs.\cite{Chandrasekaran2020PRB,Yuan2020PRB} and have universal character as function of chemical potential divided by temperature. In addition, we compare the analytic expressions with numerical calculations for several tight-binding models (modified graphene \cite{Classen2020} and Lieb lattice \cite{Shen2010PRB}) that host different types of high-order saddle points in dispersion. Then, we extend the analysis to the tight-binding model of square-octagon lattice, where the dispersion has qualitative changes of type from saddle points to completely flat bands with changing model parameters \cite{Sheng2012,Yamashita2013PRB,Oriekhov2021}. The results for logarithmic van Hove singularities presented in the main text support the qualitative conclusions of Ref.\cite{Galperin2018} that the differential entropy has characteristic dip and peak structure, and passes through zero at the saddle point level. Notably, we find that the differential entropy near high-order vHs level has even more pronounced features such as dip or peak, and its' behavior is fully defined by vHs type and temperature.    

The paper is organized as follows: in Sec.\ref{sec:entropy-calculation} we present main definitions and derive analytic expressions for differential entropy. Also we identify the distinguishing features of the entropy per particle that characterize vHs. Next in Sec.\ref{sec:models-numerical} we present the comparison of analytic expressions with numerical calculations of the differential entropy for DoS found from tight-binding models. Finally, in Sec.\ref{sec:conclusions} we discuss the obtained results and their qualitative consequences that can be observed in experiments. The technical details of analytic derivations are discussed in Appendices \ref{appendix:integrals} and \ref{appendix:epsilon-0}.

\section{Differential entropy near van Hove singularities}
\label{sec:entropy-calculation}
The differential entropy per electron in given system can be evaluated using Maxwell relation
\begin{align}\label{eq:s-Maxwell-1}
	s\equiv\left(\frac{\partial S}{\partial n}\right)_{T}=-\left(\frac{\partial \mu}{\partial T}\right)_{n}=\left(\frac{\partial n}{\partial T}\right)_{\mu}\left(\frac{\partial n}{\partial \mu}\right)_{T}^{-1},
\end{align}
where the derivative of chemical potential is expressed through derivatives of electron concentration $n(\mu, T)$. The quantity $n(\mu, T)$ itself can be expressed through the density of states in the system:
\begin{align}\label{eq:particle-concentration}
	n(\mu, T)=\int_{-\infty}^{\infty} \frac{D(\varepsilon)}{\exp \left(\frac{\varepsilon-\mu}{T}\right)+1} d \varepsilon.
\end{align}
Here we set the Boltzmann constant to be equal $k_B=1$ and measure the temperature in energy units. For the general 2D tight-binding model the density of states is defined as
	\begin{align}\label{eq:DOS-definition}
	D(\varepsilon)=\sum_{\lambda} \int \frac{d^{2} k}{(2 \pi)^{2}} \delta\left[\varepsilon-E_{\lambda}(\mathbf{k})\right],
\end{align}
where the summation accounts for the band index $\lambda$ and possible spin degrees of freedom. Substituting these quantities back to Eq.\eqref{eq:s-Maxwell-1} one finds the well-known expression for the differential entropy,
\begin{align}\label{eq:entropy-definition-integral}
	s(\mu, T)=\frac{1}{T} \frac{\int_{-\infty}^{\infty} d \varepsilon D(\varepsilon)(\varepsilon-\mu) \cosh ^{-2}\left(\frac{\varepsilon-\mu}{2 T}\right)}{\int_{-\infty}^{\infty} d \varepsilon D(\varepsilon) \cosh ^{-2}\left(\frac{\varepsilon-\mu}{2 T}\right)}.
\end{align}
This expression is used below in Sec.\ref{sec:models-numerical} for numerical calculations. 

Now let us concentrate on the specific expressions for DoS, where the energy $\epsilon$ is measured from the level of saddle point: 
\begin{align}\label{eq:dos-log}
	&D_{\log}(\epsilon)=C_1\log\frac{\epsilon_0}{|\epsilon|},\\
	\label{eq:dos-power}
	&D_{\alpha}(\epsilon)=C_2\bigg[D_{+}\Theta(\epsilon)\left(\frac{\epsilon_0}{\epsilon}\right)^{\alpha}+D_{-}\Theta(-\epsilon)\left(\frac{\epsilon_0}{|\epsilon|}\right)^{\alpha}\bigg], \\
	\label{eq:dos-flat}
	&D_{\text{flat}}(\epsilon)=C_3\,\delta(\epsilon-\epsilon_0).
\end{align}
Here the first expression with logarithmic behavior corresponds to ordinary van Hove singularity and the second expression with power law divergence works in the vicinity of high-order vHs \cite{Yuan2019Nature,Yuan2020PRB,Chandrasekaran2020PRB}. The power law exponent has natural upper limit $\alpha<1$, which comes from the requirement that the total concentration of particles in the system is finite. $\Theta(\epsilon)$ denotes the Heaviside step function and used to describe the behavior on different sides of saddle point. The numerical factors $D_{\pm}$ define whether the DoS is symmetric or asymmetric around the singularity \cite{Chandrasekaran2020PRB}, and the constants $C_1$, $C_2$ and $\epsilon_0$ are used to properly normalize the expressions and take into account for material-dependent parameters. Finally, Eq.\eqref{eq:dos-flat} describes the DoS of perfectly flat band placed at energy level $\epsilon_0$, and is typical effective approximation of DoS for many tight-binding models such as for Lieb \cite{Shen2010PRB} and dice lattices \cite{Gorbar2019PRB}.

\begin{figure}
	\centering
	\includegraphics[scale=0.45]{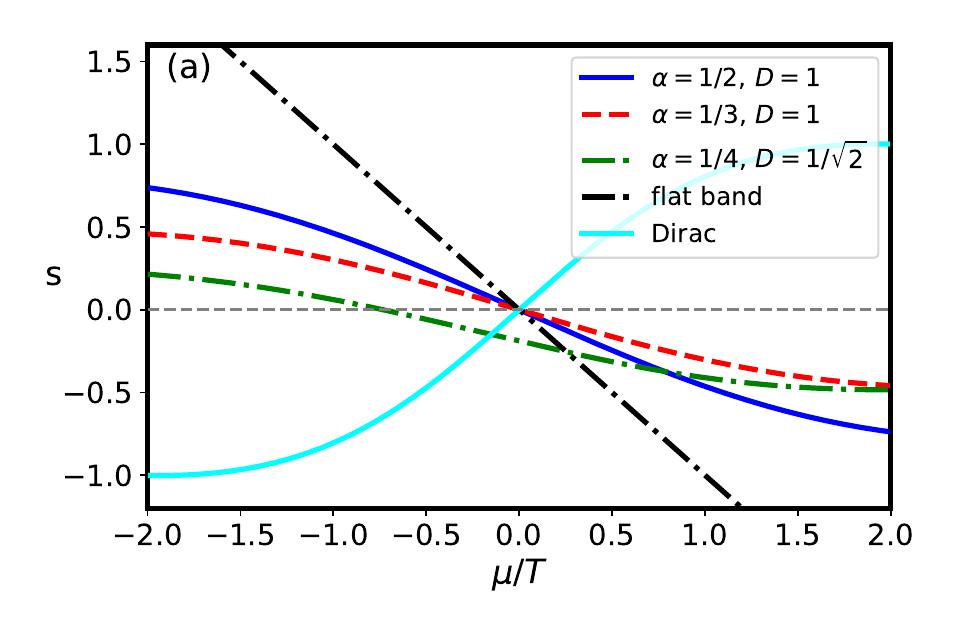}
	\includegraphics[scale=0.45]{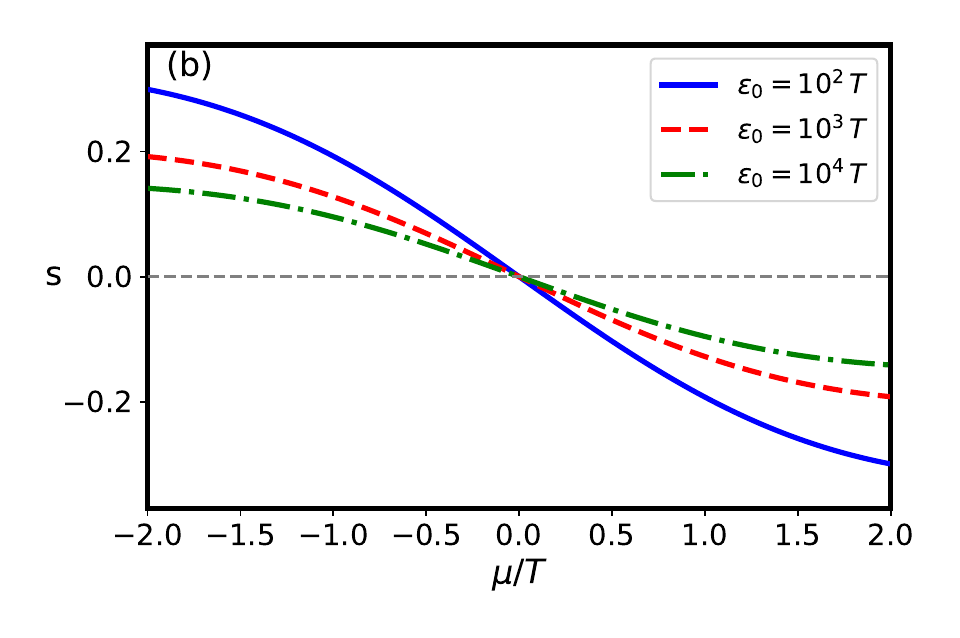}
	\caption{Panel (a): plot of exact expression for $s_{\alpha}(\mu,T)$ given by Eq.\eqref{eq:entropy-power-alpha} as a function of chemical potential normalized by temperature, $\mu/T$, for different values of parameter $\alpha$ and ratios $D=D_{+}/D_{-}$. The values are taken from Ref.\cite{Chandrasekaran2020PRB} and describe the high-order vHs that can appear in real 2D crystals. The plot range in $\mu/T$ is chosen to cover the region of fast change of Fermi function \cite{Tsaran2017Nature}, where the predicted behavior is expected to occur. The curves of the differential entropy for symmetric $D=1$ and asymmetric $D\neq 1$ high-order vHs are compared with flat band \eqref{eq:entropy-flat-band} and Dirac cone Eq.\eqref{eq:Dirac-entropy} results. The slope of $s_{Dirac}$ curve has opposite sign comparing to other presented curves. The flat band curve has the largest slope comparing to power law $s_{\alpha}(\mu,T)$ curves as predicted by Eqs.\eqref{eq:entropy-flat-band} and \eqref{eq:expansion-power-law}. Panel (b): plot of expression \eqref{eq:entropy-log} for $s_{log}(\mu,T)$ as a function of chemical potential normalized by temperature $\mu/T$ for different values of dimensional parameter $\epsilon_0$ of the model, measured in units of $T$.}
	\label{fig:power-law-entropy}
\end{figure}
 
To evaluate the differential entropy analytically we start with expression for particle concentration, Eq.\eqref{eq:particle-concentration}, and substitute expressions for DoS \eqref{eq:dos-log}, \eqref{eq:dos-power} and \eqref{eq:dos-flat}. Performing the calculation described in Appendix \ref{appendix:integrals}, we find the closed-form expression in terms of polylogarithm $\text{Li}_{a}(x)$ and gamma $\Gamma(x)$ functions \cite{Whittaker2009} for $n_{\alpha}(\mu,T)$. 
 From particle concentration $n_{\alpha}(\mu,T)$ given by Eq.\eqref{eq:vHs-particle-concentration} we find the following expression for the differential entropy
\begin{align}
	\label{eq:entropy-power-alpha}
	s_{\alpha}(\mu,T)=(\alpha-1)\frac{D_{-}\mathrm{Li}_{1-\alpha}\left(-z\right)-D_{+}\mathrm{Li}_{1-\alpha}\left(-\frac{1}{z}\right)}{D_{-}\mathrm{Li}_{-\alpha}\left(-z\right)+D_{+}\mathrm{Li}_{-\alpha}\left(-\frac{1}{z}\right) }-\frac{\mu}{T},\quad z=\exp\left(-\frac{\mu}{T}\right).
\end{align}  
From this expression it is evident that differential entropy for power-law divergent DoS depends only on dimensionless combination $\mu/T$. Thus, the function $s_{\alpha}(\mu,T)$ is universal for different classes of materials, some of which are listed in classifications from Refs.\cite{Yuan2020PRB,Chandrasekaran2020PRB}.  Notably, the Eq.\eqref{eq:entropy-power-alpha} can be also used for the differential entropy near Dirac cone spectrum. Setting $\alpha=-1$,  $D_{+}=D_{-}$ and using the identity $\text{Li}_{1}(z)=-\log(1-z)$, we find the following expression, which agrees with results of Ref.\cite{Tsaran2017Nature}:
\begin{align}\label{eq:Dirac-entropy}
	s_{Dirac}(\mu,T)=\frac{\mathrm{Li}_{2}\left(-z\right)-\mathrm{Li}_{2}\left(-\frac{1}{z}\right)}{\log\left[2\cosh \left(\frac{\mu}{2T}\right)\right]}-\frac{\mu}{T}.
\end{align}
The differential entropy function $s_{\alpha}(\mu,T)$ is plotted in Fig.\ref{fig:power-law-entropy} for several values of $\alpha$ and corresponding examples of $D_{+}$ and $D_{-}$ ratios, and compared with $s_{Dirac}(\mu,T)$. The qualitative difference between two cases is the different sign of the slope of the curves as functions of $\mu/T$ when $s(\mu,T)$ passes through zero. Thus one can distinguish van Hove singularities from Dirac cones in the characteristic dip and peak structure of the differential entropy using the sign of the curve slope.

For the logarithmic van Hove singularity we find the following expression for differential entropy (see Appendix \ref{appendix:integrals}):
\begin{align}
	\label{eq:entropy-log}
	s_{log}(\mu, T)=&-\left[\frac{\mu}{T}\left(\mathrm{Li}_{0}^{(1,0)}(-z)+\mathrm{Li}_{0}^{(1,0)}\left(-\frac{1}{z}\right)+1\right)+\mathrm{Li}_{1}^{(1,0)}(-z)-\mathrm{Li}_{1}^{(1,0)}\left(-\frac{1}{z}\right)\right]\nn
	&\times\left[\mathrm{Li}_{0}^{(1,0)}(-z)+\mathrm{Li}_{0}^{(1,0)}\left(-\frac{1}{z}\right)-\log \left(\frac{\varepsilon_{0}}{T}\right)+\gamma\right]^{-1}.
\end{align}
Here $\gamma\approx 0.5772$ is the Euler constant, and the upper indexes in polylogarithm functions correspond to the order of derivative with respect to argument $\text{Li}^{(n,m)}_{a}(x)\equiv \p_a^n \p_x^m \text{Li}_{a}(x)$. 
In this case the dependence of $s_{log}(\mu, T)$ on $\mu$ and $T$ does not reduce to the single dimensionless variable $\mu/T$ (as it was for power-law-diverging DoS,  Eq.\eqref{eq:entropy-power-alpha}), since the logarithms in the denominator bracket contains also $\epsilon_0/T$. In the tight-binding models the parameter $\epsilon_{0}$ usually appears as combination of spectrum decomposition coefficient $a,b$ in dispersion $\epsilon\approx const-a k_x^2+b k_y^2$, and momenta cut-off parameter that defines the applicability range of the model \cite{Yuan2019Nature}. Thus, the obtained formula for $s_{log}(\mu,T)$ has less universal character, as one should always take into account the material-specific constant. Assuming that $\epsilon_{0}$ is of the same order as tight-binding parameters $\epsilon_{0}\sim 1\,\text{eV}$, and characteristic temperatures in experiment are usually from $<1$ up to 100 K, one can estimate the ratio $\epsilon_{0}/T\approx 10^2 \,- 10^4$. We plot $s_{log}(\mu,T)$ in the panel (b) of Fig.\ref{fig:power-law-entropy} to compare it with flat band $s_{flat}$ and power law $s_{\alpha}$ differential entropy expressions.

%\begin{figure}
%	\centering
%	\includegraphics[scale=0.62]{entropy_analytic_log_exact.pdf}
%	\includegraphics[scale=0.62]{entropy_analytic_log_series.pdf}
%	\caption{Left panel: plot of exact expression for $s_{log}(\mu,T)$ as a function of chemical potential normalized by temperature $\mu/T$ for different values of dimensional parameter $\epsilon_0$ of the model. Right panel: series decomposition up to linear order in $\mu/T$.}
%	\label{fig:logarithm-entropy}
%\end{figure}
Finally, in the case of fully flat band as the limiting case of saddle points in dispersion the DoS is given by Eq.\eqref{eq:dos-flat}, where $\epsilon_0$ is the flat band level. The expression for the entropy per particle in the vicinity of $\mu=\epsilon_0$ takes very simple form due to integration with delta-function in Eq.\eqref{eq:entropy-definition-integral}:
\begin{align}\label{eq:entropy-flat-band}
	s_{flat\,band}(\mu, T)&%=\frac{1}{T} \frac{\int_{-\infty}^{\infty} d \varepsilon \delta(\varepsilon-\varepsilon_0)(\varepsilon-\mu) \cosh ^{-2}\left(\frac{\varepsilon-\mu}{2 T}\right)}{\int_{-\infty}^{\infty} d \varepsilon \delta(\varepsilon-\varepsilon_0) \cosh ^{-2}\left(\frac{\varepsilon-\mu}{2 T}\right)}
	=\frac{\varepsilon_0-\mu}{T},
\end{align}
and shows that the entropy decreases linearly with chemical potential passing through zero at $\mu=\epsilon_0$.
To compare the flat band case with  previously-derived formulas for differential entropy, we expand Eqs.\eqref{eq:entropy-power-alpha} and \eqref{eq:entropy-log} into series around $\mu/T=0$ up to first order. For the power-law vHs the expansion is the following:
\begin{align}\label{eq:expansion-power-law}
	&s_{\alpha}(\mu,T)\approx-\frac{\left(2^{\alpha}-1\right)(\alpha-1)(D-1) \zeta(1-\alpha)}{\left(2^{\alpha+1}-1\right)(D+1) \zeta(-\alpha)}-\alpha\frac{\mu}{T}+(\alpha-1)\frac{\left(2^{\alpha}-1\right)\left(2^{\alpha+2}-1\right)(D-1)^{2} \zeta(-\alpha-1) \zeta(1-\alpha)}{\left(2^{\alpha+1}-1\right)^{2}(D+1)^{2} \zeta(-\alpha)^{2}}\frac{\mu}{T}.
\end{align}
Here we denote the radio of numerical coefficients $D_{\pm}$ as $D=D_{+}/D_{-}$ and $\zeta(x)$ denotes the Riemann zeta-function. 
Note that the constant term vanishes for symmetric saddle point dispersion, while the linear term reduces to $-\alpha \frac{\mu}{T}$. Thus, the slope of entropy curve plotted as a function of chemical potential immediately gives the divergence exponent and van Hove singularity type. Also, the physical upper limit for $\alpha$ is $\alpha<1$ and the $\alpha=1$ corresponds to exactly flat band \eqref{eq:entropy-flat-band}. Thus, the entropy per particle clearly measures the band "flatness". 

For the logarithmic vHs we have the following expansion up to linear order in $\mu/T$:
\begin{align}
		&s_{log}(\mu,T)\approx 
		-\frac{1}{\log \left(\frac{\epsilon_0}{T}\right)+\gamma +\log \left(\frac{2}{\pi }\right)}\frac{\mu}{T}.
\end{align}
Notably, the differential entropy $s_{log}(\mu,T)$ always passes through zero when $\mu$ passes through the energy level of saddle point. The prefactor contains logarithmic dependence on $\epsilon_{0}$ parameter and in principle allows one to estimate the value of this parameter for given model from the differential entropy curve slope. 

\section{Differential entropy as probe of vHs type in tight-binding models}
\label{sec:models-numerical}
\begin{figure}
	\centering
	\includegraphics[scale=0.52]{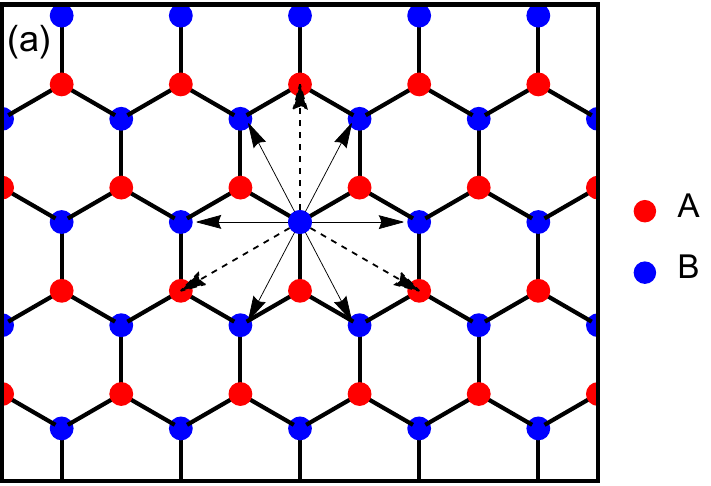}
	\includegraphics[scale=0.62]{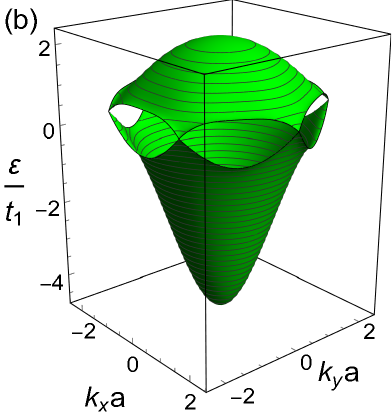}\quad
	\includegraphics[scale=0.37]{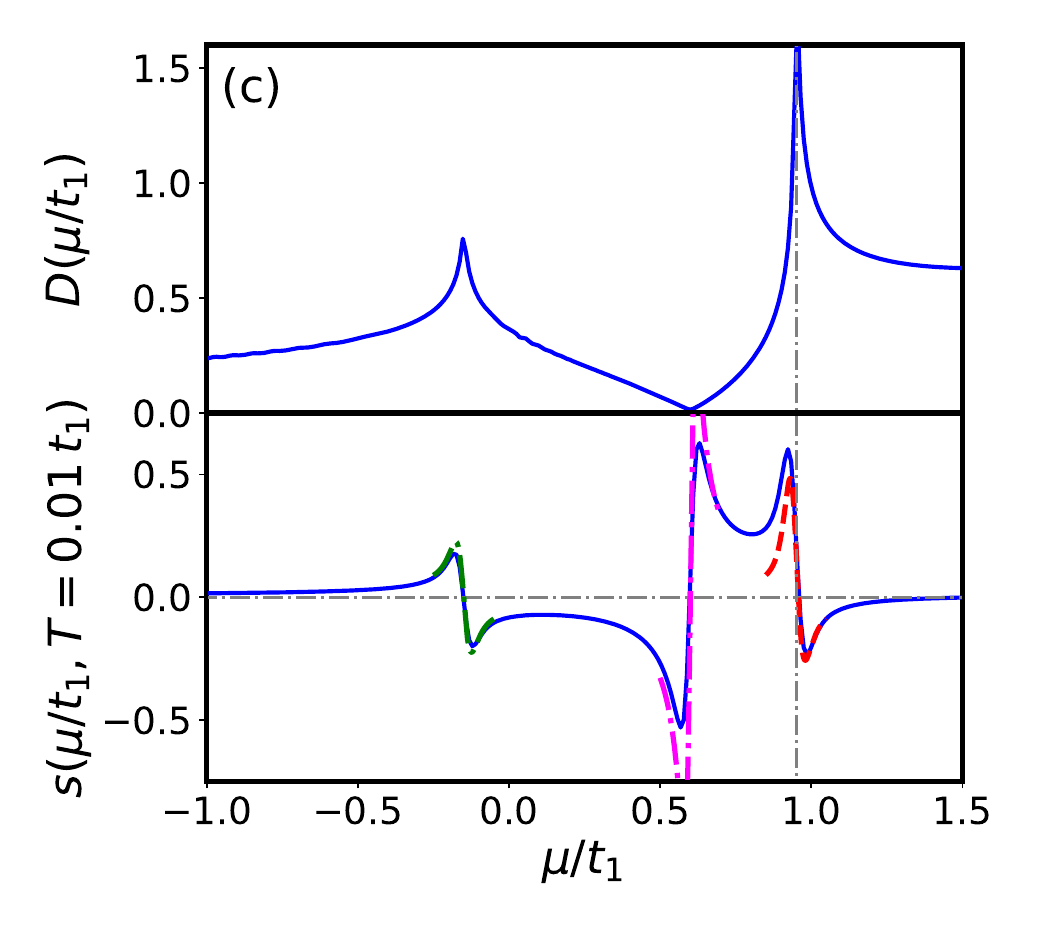}
	\caption{Panel (a): the geometry of graphene lattice with nearest neighbor (solid lines), next-nearest neighbor (solid arrows) and next-next-nearest neighbor (dashed arrows)  hopping parameters. Panel (b): spectrum of the model for the critical values of tight-binding parameters $t_2=0.2\, t_1$ and $t_3=0.15\, t_1$. The high-order saddle points are placed on the boundaries of BZ in the upper band and have very flat dispersion along the boundary (can be inferred from constant energy curves). Panel (c): the upper plot shows the numerically calculated DoS for spectrum from (b). The plot range covers the regions around both van Hove singularities. Usual and high-order van Hove singularities are visible as high narrow peaks. The lower plot demonstrates the numerically calculated entropy per particle for temperature $T=0.01 t_1$ and plotted in the same range of chemical potentials. The analytic predictions for high-order vHs at $E=0.95 t_1$, for logarithmic vHs at $E=-0.15 t_1$ and for Dirac cone at $R=0.6 t_1$ are shown as red \eqref{eq:entropy-power-alpha}, green \eqref{eq:entropy-log} and pink \eqref{eq:Dirac-entropy} dashed lines respectively and demonstrate good agreement with numerical results.}
	\label{fig:graphene-lattice}
\end{figure}
In the present section we analyze differential entropy per particle for several lattice models hosting regular and high-order van Hove singularities, and flat bands. To analyze different van Hove singularities we firstly use the toy model from Ref.\cite{Classen2020} of monolayer graphene with tunable next-nearest neighbor and next-next-nearest neighbor hopping parameters. In this model one can switch between usual and high-order van Hove singularities by adjusting these parameters to specific values. Using this example, we study asymmetric high-order vHs with divergence exponent of DoS $\alpha=1/4$.  Next we take the pseudospin-1 tight-binding model which hosts ideal flat band - Lieb lattice. In this model one can analyze to which extend the flat band solely determines the differential entropy behavior as strongly dominating feature in density of states. Finally, we analyze the tight-binding model of square-octagon lattice, in which the most asymmetric type of high-order vHs is realized with one of $D_{\pm}$ coefficients equal to zero and $\alpha=1/2$. Also, by tuning the hopping parameters one can find two flat bands in this model, and we also briefly discuss such setting.

\subsection{Graphene model with next- and next-next-nearest neighbor hopping parameters}
\label{sec:graphene}

Let us start with the simple model of graphene from Ref.\cite{Classen2020} that realizes both usual and high-order vHs at different values of hopping parameters. The lattice is schematically shown in Fig.\ref{fig:graphene-lattice} (a). The tight-binding Hamiltonian in the basis of A, B sublattice components $\psi=(\psi_A,\,\psi_B)^{T}$ has the form:
\begin{align}\label{eq:h-graphene-NNN}
	&	H_{graphene}=\begin{pmatrix}
		-t_2 \beta(\vec{k}) & t_1 \alpha(\vec{k})+t_3\gamma(\vec{k})\\
		t_1 \alpha^{*}(\vec{k})+t_3\gamma^{*}(\vec{k}) & -t_2 \beta(\vec{k})
	\end{pmatrix},
\end{align}
where the phase sums over nth-nearest neighbors are given by
\begin{align}
		&\alpha(\vec{k})=\sum_{n=1}^{3} e^{-i \vec{k} \cdot \vec{a}_{n}}, \quad\beta(\vec{k})=\sum_{n=1}^{6} e^{-i \vec{k} \cdot \vec{b}_{n}},\quad \gamma(\vec{k})=\sum_{n=1}^{3} e^{-i \vec{k} \cdot \vec{c}_{n}}.
\end{align}
The corresponding sets of vectors that correspond to different neighboring atoms are defined as 
\begin{align}
	&\vec{a}_{1}=(\sqrt{3}, 1) / 2, \quad\vec{a}_{2}=(-\sqrt{3}, 1) / 2, \quad \vec{a}_{3}=(0,-1)\nn
\end{align}
for the first neighbors,
\begin{align} 
	&\vec{b}_{1}=(\sqrt{3}, 0),\quad \vec{b}_{2}=(\sqrt{3}, 3) / 2,\quad \vec{b}_{3}=(-\sqrt{3}, 3) / 2,\quad \vec{b}_{4}=-\vec{b}_{1},\quad \vec{b}_{5}=-\vec{b}_{2},\quad \vec{b}_{6}=-\vec{b}_{3}
\end{align}
for the second neighbors and 
\begin{align}
	\vec{c}_{1}=-2 \vec{a}_{1},\quad \vec{c}_{2}=-2 \vec{a}_{2}, \quad \vec{c}_{3}=-2 \vec{a}_{3}
\end{align}
for the third. These vectors are shown in Fig.\ref{fig:graphene-lattice}(a) as solid lines, solid and dashed arrows, respectively. The hopping parameters $t_i$ correspond to $i^{th}$-nearest neighbor contributions. 
The highly-symmetric points of this model are $K$-point in the corners of hexagonal Brillouin zone, which are given by:
\begin{align}
	K_{1}=2 \pi / 3(1 / \sqrt{3}, 1), \quad K_{2}=2 \pi / 3(-1 / \sqrt{3}, 1),
\end{align}
where one finds usual Dirac cones, and $M$-points in the middles of BZ edges:
\begin{align} 	
	 M_{1}=\pi(0,2 / 3),\quad M_{2}=\pi(-1 / \sqrt{3}, 1 / 3), \quad M_{3}=-\pi(1 / \sqrt{3}, 1 / 3),
\end{align}
where the saddle points corresponding to van Hove singularities are placed. Diagonalizing the Hamiltonian, one finds the two bands, that are described by the following dispersion relation:
\begin{align}\label{eq:graphene-NNN-spectrum}
	\epsilon_{\pm}(\vec{k})=\pm\left|t_{1} \alpha(\vec{k})+t_{3} \gamma(\vec{k})\right|-t_{2} \beta(\vec{k}).
\end{align} 
The main feature of this model is that at special values of the 3rd-NN parameters 
\begin{align}
	t_{3} \rightarrow t_{3, c}=\left(t_{1}-2 t_{2}\right) / 4
\end{align}
one meets the high-order van Hove singularity instead of classical one \cite{Classen2020}. The bands $\epsilon_{\pm}(\vec{k})$ are plotted in Fig.\ref{fig:graphene-lattice} specifically for this relations of parameters ($t_2=0.2 t_1,\,\,t_3=0.15 t_2$). 
The corresponding density of states near saddle point in upper band is approximately given by
\begin{align}
	D(\epsilon)\approx\rho \begin{cases}D_{+} \epsilon^{-1 / 4} & \text { for } \epsilon>0 \\ D_{-}|\epsilon|^{-1 / 4} & \text { for } \epsilon<0\end{cases}
\end{align}
with $D_{-}=D_{+}/\sqrt{2}$ and $\rho$ is the proper dimensional parameter. The high-order vHs appears only in upper band, while in the lower band it is always usual logarithmic vHs. We calculate its exact positions from $\epsilon_{\pm}(\vec{k})$, taking $M_1$ point:
\begin{align}\label{eq:vHs-up}
	\epsilon_{+}(\vec{k}=\vec{M}_1)=|t_1-3 t_3| + 2 t_2,\quad \epsilon_{-}(\vec{k}=\vec{M}_1)=-|t_1-3 t_3| + 2 t_2..
\end{align}

In the upper plot in Fig.\ref{fig:graphene-lattice}(c) we plot the density of states for the tight-binding model \eqref{eq:h-graphene-NNN}. The delta-functions in the density of states integrals \eqref{eq:DOS-definition} were regularized by introducing finite energy level width in the form of Lorentzian shape with parameter $\Gamma$:
\begin{align}\label{eq:delta-regularize}
	\delta(\epsilon-E_{\lambda})\to \frac{1}{\pi}\frac{\Gamma}{
	\Gamma^2+(\epsilon-E_{\lambda})^2}.
\end{align}
The ``width" $\Gamma$ is taken to be much smaller that characteristic energy scale defined by hopping parameter $t$ and smaller that $T$ to avoid important features from being hidden. Such level broadening can occur due to the presence of disorder, boundaries, defects and electron-electron, electron-phonon interactions in real systems. The experiments suggest that $\Gamma$ can be of the order of 10 meV \cite{Andrei2012,Xie2019Nature}. Recently the more detailed theoretical discussion of the role of disorder on DoS smearing was given in Ref.\cite{Chandrasekaran2022} using Born approximations. Here for the numerical calculations we took either $\Gamma=0.005 t_1$ or $\Gamma=0.01 t_1$ to check that results do not strongly depend on choice as long as $\Gamma<T$ (room temperature is around $T\sim 20 meV$ in energy units). The integrals are evaluated numerically using equally-spaced grid with small enough discretization steps in $k_x$, $k_y$ and $\epsilon$. To evaluate DoS, we calculate the spectrum in each point $(k_{x},k_y)$ of discretization inside first BZ and perform summation over bands using smeared delta-functions \eqref{eq:delta-regularize}. The energy integrals for the differential entropy (see Eq.\eqref{eq:entropy-definition-integral}) are bounded on both sides with cut-off parameter that is larger than the total band width of tight-binding model. This ensures that the contribution of all bands are included into final result.

  In the lower plot in Fig.\ref{fig:graphene-lattice}(c) we plot the numerically calculated differential entropy as a function of chemical potential for $T=0.01 t_1$. The interval of both upper and lower parts is selected such that it covers all characteristic features from spectrum in panel (b). The analytic curve for $s_{\alpha}(\mu,T)$ from Eq.\eqref{eq:entropy-power-alpha} is plotted around chemical potential level $\epsilon_{+}(\vec{k}=\vec{M}_1)$ with $\alpha=1/4$ and $D_{+}/D_{-}=\sqrt{2}$ as red dashed line. It demonstrates good agreement with numerically calculated entropy per particle in the vicinity of vHs level and predicts the slope of the curve that passes through zero. Away from that level it also describes the peak and reduction of differential entropy, but the precision is not so accurate. As was noted in Refs.\cite{Tsaran2017Nature,Galperin2018}, one should expect the manifestation of main features in entropy in $-T,T$ interval around the corresponding energy level. In our case the theoretical curve is plotted in the range $-10T, 10T$, and still remains mostly valid. On the smaller chemical potential the analytic formula predicts less pronounced peak than the one that actually appears. The possible reason of such underestimation is the fact that the closely-placed Dirac points (see panel (b)) induce the entropy spikes \cite{Tsaran2017Nature}, and these spikes are partially merged with that coming from vHs. Notably, the Eq.\eqref{eq:Dirac-entropy} that describes differential entropy per particle near Dirac cone works well predicting the slope and its sign, but instead overestimates the height of dip and peak structure. The latter may be caused by level broadening as DoS does not drop to zero at Dirac cone if $\Gamma\neq 0$ \cite{Tsaran2017Nature}.

\begin{figure}
	\centering
	\includegraphics[scale=0.5]{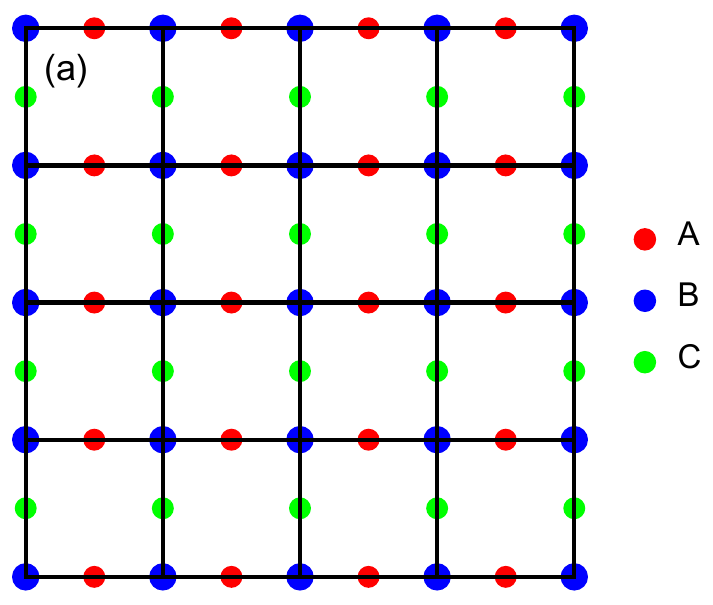}\qquad
	\includegraphics[scale=0.5]{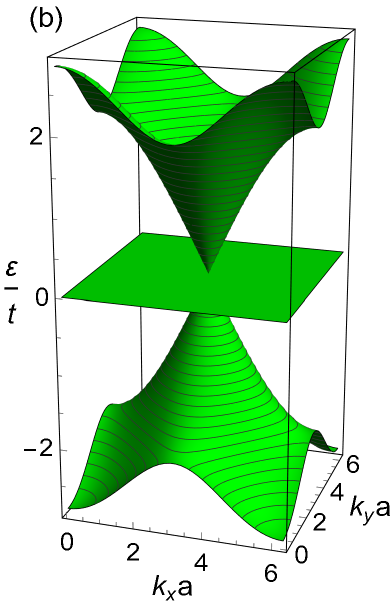}\qquad 
	\includegraphics[scale=0.5]{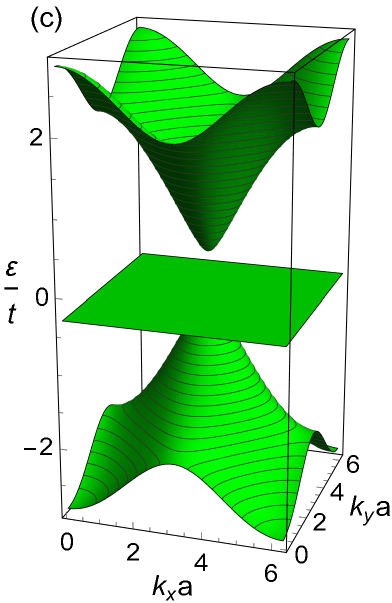}
	\caption{Panel (a): the geometry of Lieb lattice. Three sublattices are shown with different colors and only hopping parameters between A-B and B-C are included into tight-binding Hamiltonian \eqref{eq:h-Lieb}. Panels (b,c): spectrum of the tight-binding Lieb model, defined via Eq.\eqref{eq:bands-Lieb} for gap parameters $\Delta=0$ and $\Delta=0.3 t$, respectively.}
	\label{fig:Lieb-lattice}
\end{figure}

  Finally, we analyze how the formula for differential entropy \eqref{eq:entropy-log} works around logarithmic vHs. The panel (b) in Fig.\ref{fig:power-law-entropy} suggests that expected behavior is similar to one observed in the vicinity of usual vHs level in Fig.\ref{fig:graphene-lattice} (c). There is some arbitrariness in choosing the $\epsilon_0$ parameter as it usually contains momentum cut-off \cite{Yuan2019Nature}. It is relatively hard to take it from tight-binding calculation of DoS, since there are other contributions and it is difficult to compare logarithms. However, this can be found by comparing slope of the differential entropy curve with theoretic prediction. In the panel (c)
  we plotted the analytic curve given by Eq.\eqref{eq:entropy-log} with $\epsilon_{0}=5t_1$ as green dashed line. One should note that the parameter $\epsilon_{0}$ does not have strict physical limits as it comes as normalization under logarithm, and appears as combination of constants in band decomposition and cut-off parameters over momenta in effective model. We checked numerically that $\epsilon_0$ is of the order of $10 t_1$ (thus, $\epsilon_0>100 \,\,T$). In Appendix \ref{appendix:epsilon-0} we performed estimation of $\epsilon_0$ using more precise series expansion of dispersion around $\vec{M}_1$ point, and found $\epsilon_0 \approx 13 t_1$. This is in good agreement with numerical calculation. Also this result together with Fig.\ref{fig:power-law-entropy} shows that the slope of $s_{log}$ function is always smaller that for $s_{\alpha}$ at vHs chemical potential. However, we should again underline that the formula \eqref{eq:entropy-log} has much less predictive power due to its non-universal character with  $\epsilon_0$ parameter inside. 
Below we analyze other systems to test the applicability of Eqs.\eqref{eq:entropy-power-alpha} and \eqref{eq:entropy-flat-band} in different cases.

\subsection{Flat band system: Lieb lattice}
\begin{figure}
	\centering
	\includegraphics[scale=0.38]{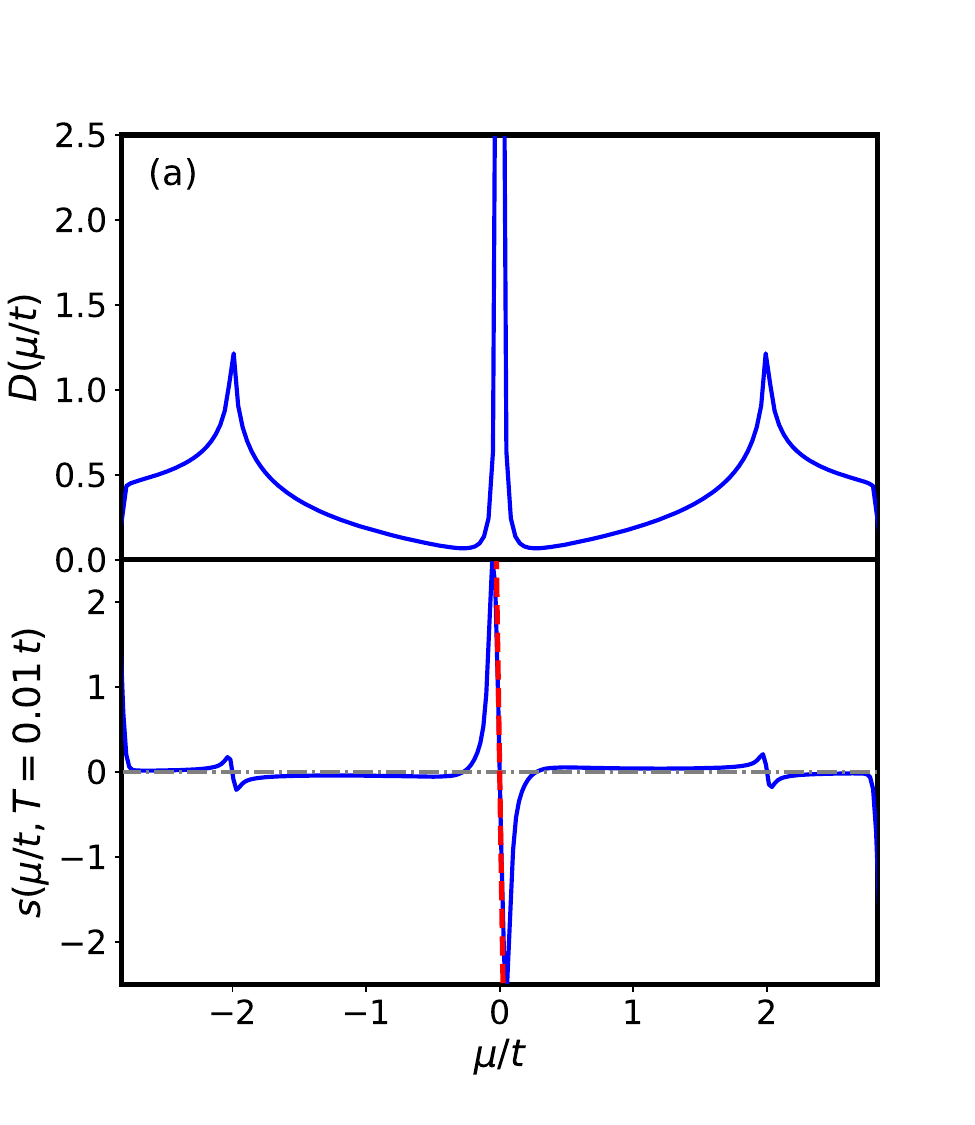}\qquad\qquad
	\includegraphics[scale=0.38]{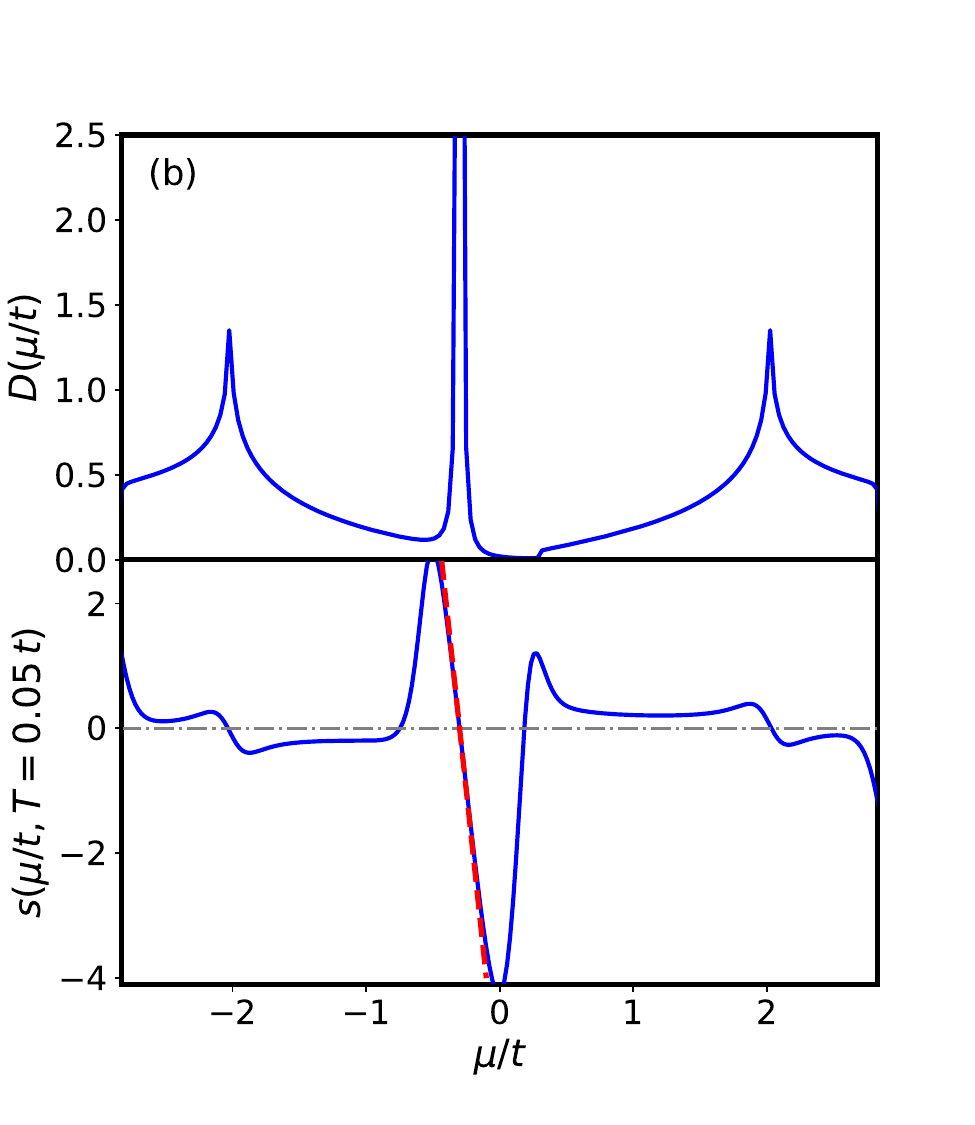}
	\caption{The correspondence between DoS on upper plots and differential entropy per particle on lower plots for the Lieb lattice model \eqref{eq:h-Lieb} as functions of chemical potential measured in units of hopping parameter $t$. Panel (a) shows the zero gap $\Delta=0$ regime for $T=0.01 t$ where the flat band meets with two Dirac cones at triply degenerate point. In panel (b) we take $\Delta=0.3\, t$ with higher temperature $T=0.05 t$. In both cases the level broadening is $\Gamma=0.005\, t$. Red dashed curves show the analytic expressions \eqref{eq:entropy-flat-band} for differential entropy near flat band level and demonstrate the good agreement with numerical results.}
	\label{fig:Lieb-lattice-entropy}
\end{figure}
The Lieb lattice is schematically shown in panel (a) of Fig.\ref{fig:Lieb-lattice}. It consists of three square sublattices, with atoms placed in the corners and in the middle of each side of elementary cell. The corresponding tight-binding Hamiltonian was described in Ref.\cite{Shen2010PRB}:
\begin{align}\label{eq:h-Lieb}
	H_{0}=\left(\begin{array}{ccc}
		-\Delta & 2 t \cos \left(k_{x} a / 2\right) & 0 \\
		2 t \cos \left(k_{x} a / 2\right) & \Delta & 2 t \cos \left(k_{y} a / 2\right) \\
		0 & 2 t \cos \left(k_{y} a / 2\right) & -\Delta
	\end{array}\right).
\end{align} 
It contains three bands, one of which is completely flat:
\begin{align}\label{eq:bands-Lieb}
	\epsilon_{\pm}(\vec{k})=\pm \sqrt{\Delta^{2}+4 t^{2}\left[\cos ^{2}\left(k_{x} a / 2\right)+\cos ^{2}\left(k_{y} a / 2\right)\right]},\quad \epsilon_{flat}=-\Delta.
\end{align}
This model was experimentally realized in optical \cite{Mukherjee2015} and electronic lattices \cite{Slot2017}. 

 In the analysis of Lieb lattice we concentrate on the differential entropy near the flat band level. In the gapless case the two Dirac cones touch flat band in the triply degenerate point $\vec{k}=(\pi/a,\pi/a)$. Thus, one would expect the appearance of spikes in differential entropy near level $\mu=0$. Also, the analytic expression \eqref{eq:entropy-flat-band} predicts that $s(\mu,T)$ should pass through zero at $\mu=0$ and the slope is defined only by inverse temperature $1/T$. 
 
 The panels (a) and (b) in Fig.
 \ref{fig:Lieb-lattice-entropy} shows 
 the numerically calculated density of states in upper plots (the level broadening $\Gamma=0.005t$ is taken to regularize delta-functions, see Eq.\eqref{eq:delta-regularize}) and entropy per particle in lower plots for the same range of chemical potentials. The difference between two panels in the presence of gap in panel (b) $\Delta=0.3 t$ and the temperatures ((a) - $T=0.01 t$, (b) - $T=0.05 t$).
 We compare the analytic expression for the entropy near flat band (red dashed curves) with the numerically calculated one. Both panels show good agreement between analytic predictions and numerical results. Note that despite the presence of Dirac cones that touch flat band from both sides in gapless case $\Delta=0$ \cite{Shen2010PRB}, the flat band solely determines the entropy behavior in surrounding interval of chemical potentials due to the diverging density of states. Thus, the obtained results should also work for other flat-band systems such as dice lattice, where the low-energy density of states has contributions from flat band and Dirac cones \cite{Raoux2014,Gorbar2019PRB}, and the flat band is expected to be stable with respect to different perturbations \cite{Oriekhov2018LTP,Dey2018PRB,Iurov2019PRB}. 

\subsection{Transition from high-order vHs to flat band: square-octagon lattice model}
\label{sec:t-graphene}
\begin{figure}
	\centering
	\includegraphics[scale=0.65]{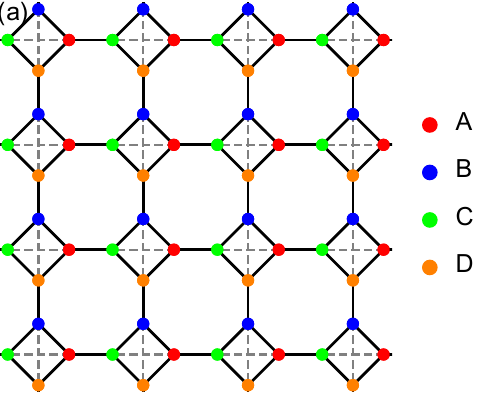}
	\includegraphics[scale=0.55]{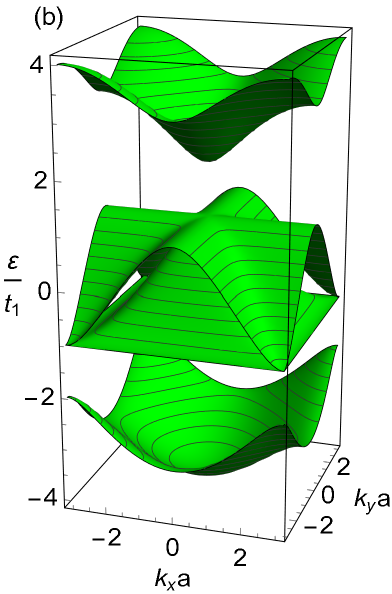}
	\includegraphics[scale=0.55]{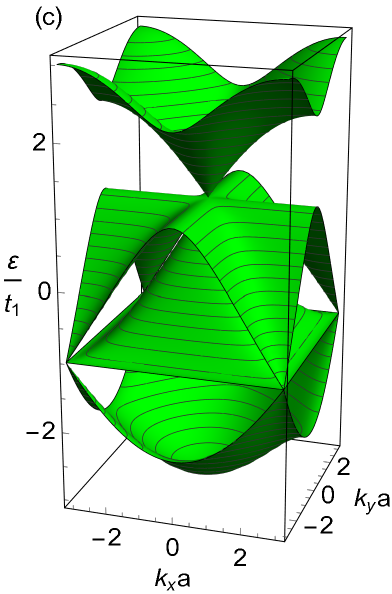}
	\includegraphics[scale=0.55]{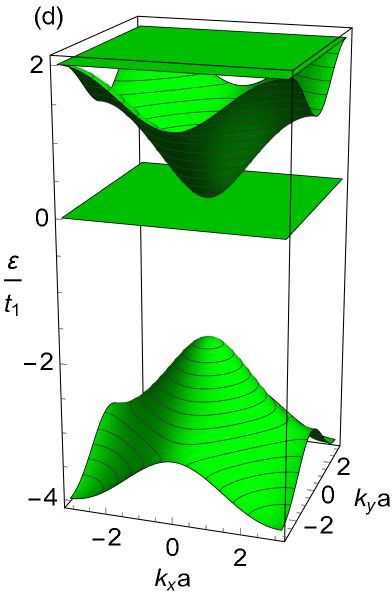}
	\caption{Panel (a): the geometry of square-octagon  lattice. Four different sublattices A, B, C and D are marked with different colors. The nearest neighbor hopping parameters $t_1$ (between A-C, B-D) and $t_2$ (edges of small squares) are shown as solid lines, and next-nearest-neighbor hopping parameters $\tau$ - as dashed lines inside small squares. In panels (b), (c) and (d) we show the three examples of spectrum for square-octagon lattice model plotted in the full BZ. The values of tight-binding hopping parameters are: (b) $t_2=\frac{3}{2}t_1$ and $\tau=0$, (c) $t_2=t_1$ and $\tau=0$, and (d) $t_2=t_1$ and $\tau=-t_1$. Panel (b) demonstrates the example of gapped model, panel (c) - the example when three-band-touching point appear. Both examples support high-order vHs that appear from flat lines in spectrum at $\epsilon=\pm t_1$. Panel (d) shows the critical example of next-nearest neighbor hopping parameter when the two bands become completely flat and touch the middle band between from both sides.}
	\label{fig:Tgraphene-lattice}
\end{figure}
Next we analyze the tight-binding model which is based on square-octagon lattice \cite{Sheng2012,Yamashita2013PRB}. This lattice attracted attention due to the possibility of existence of stable graphene allotrope called T-graphene \cite{Umemoto2010,Liu2012PRL,QinyanGu:97401}, stable $MX_2$ ($M$ =
Mo, W; $X$ = S, Se, Te) isomers \cite{Sun2015} and $\mathrm{Zn}_{2} \mathrm{O}_{2}$, $\mathrm{Zn}_{4} \mathrm{O}_{4}$ monolayers \cite{Gaikwad2017}.  It is schematically shown in panel (a) of Fig.\ref{fig:Tgraphene-lattice}. 
 As was shown in the literature, besides usual van-Hove singularities the tight-binding model of square octagon lattice always hosts high-order van Hove singularities \cite{Oriekhov2021} and in special case of fine-tuned next-nearest-neighbor parameters - two exactly flat bands \cite{Nunes2020}. The Hamiltonian, that describes the square-octagon lattice with nearest neighbor (NN) $t_{1,2}$ hopping parameters - between and inside small squares, and next nearest neighbor (NNN)  hopping parameters $\tau$ - diagonals inside small squares, has the form:
\begin{align}
	H_{T g}(\boldsymbol{k})=-\left(\begin{array}{cccc}
		0 & t_{2} & t_{1} e^{i k_{x} a}+\tau & t_{2} \\
		t_{2} & 0 & t_{2} & t_{1} e^{i k_{y} a}+\tau \\
		t_{1} e^{-i k_{x} a}+\tau & t_{2} & 0 & t_{2} \\
		t_{2} & t_{1} e^{-i k_{y} a} +\tau& t_{2} & 0
	\end{array}\right).
\end{align}
The spectrum defined by this Hamiltonian contains four bands. For arbitrary values of hopping parameters the exact expressions for energy bands are defined by fourth-order equation \cite{Sheng2012,Yamashita2013PRB,Oriekhov2021} and are very complicated. However, the expansions of spectrum near saddle points are relatively easy to obtain. In Fig.\ref{fig:Tgraphene-lattice} we plot several examples of spectrum for $\tau=0$ with two different ratios of $t_2/t_1$: $t_2/t_1=3/2$ in panel (b) and $t_2/t_1=1$ in panel (c), and for $\tau=-t_1$ with $t_2/t_1=1$ in panel (d), where the two flat band appear. As was noted in Ref.\cite{Oriekhov2021}, the high-order van Hove singularities appear on the levels of flat lines $\epsilon=\pm t_1$ for all values of $t_2/t_1$ if $\tau=0$. 
The position of flat lines can be shifted by nonzero $\tau$ hopping parameter, and for several critical values of $\tau$ the two bands become completely flat \cite{Nunes2020}. 

\begin{figure}
	\centering
	\includegraphics[scale=0.35]{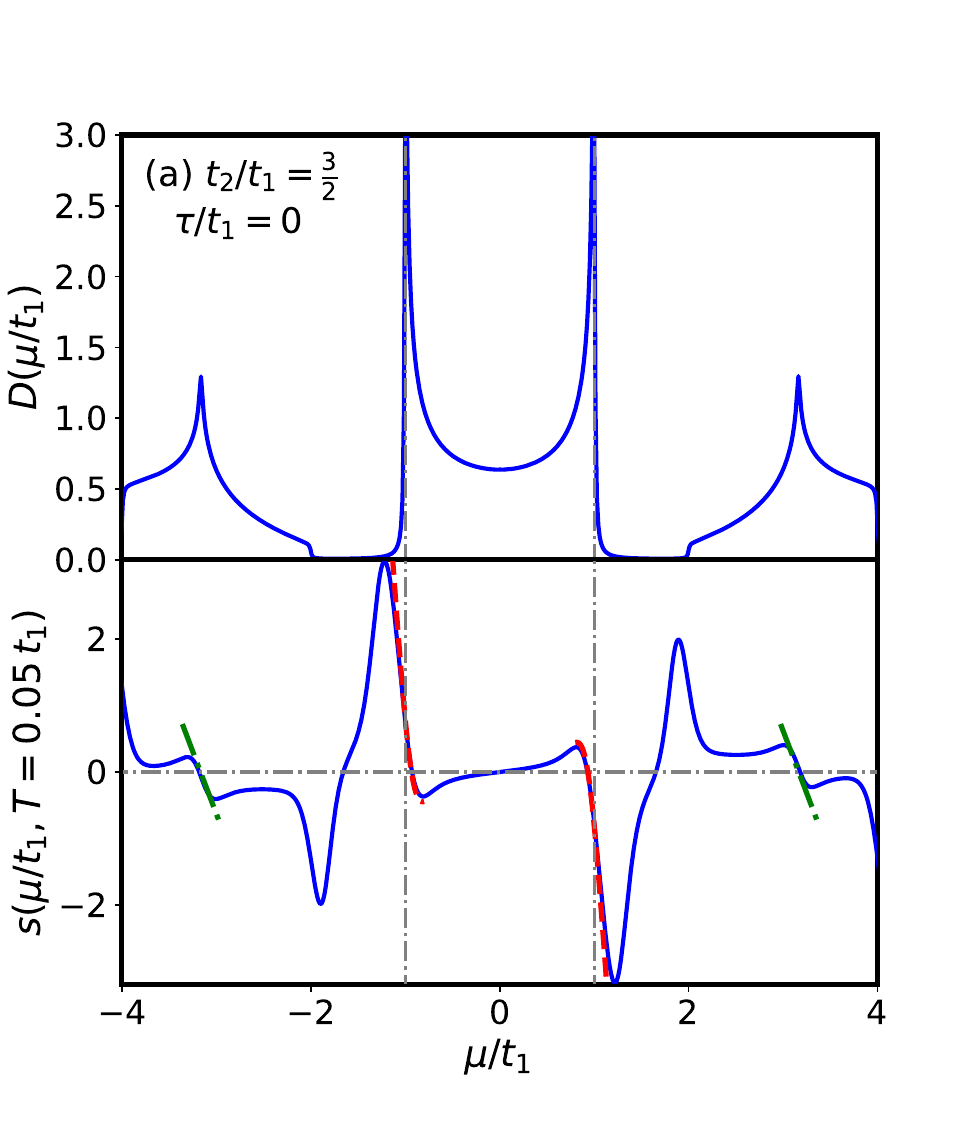}
	\includegraphics[scale=0.355]{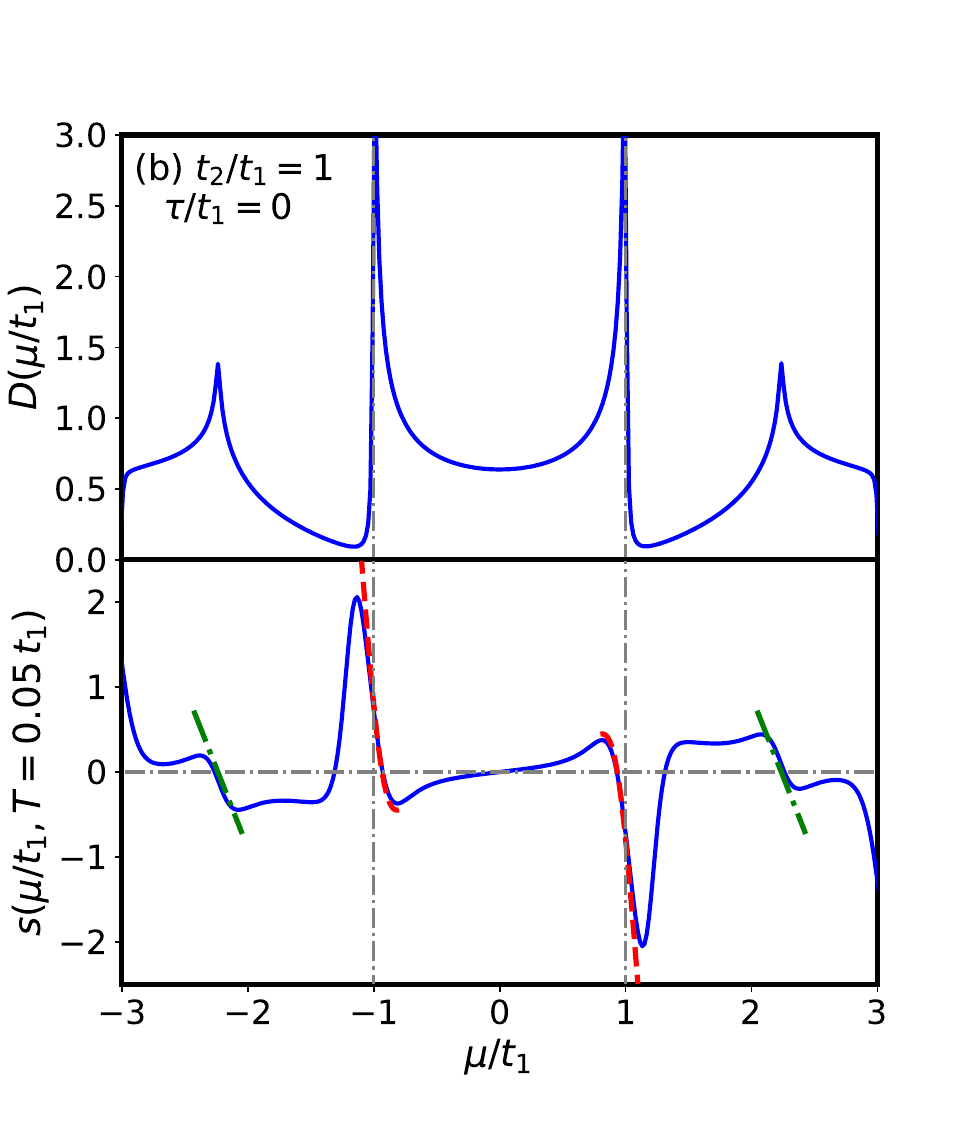}
	\includegraphics[scale=0.355]{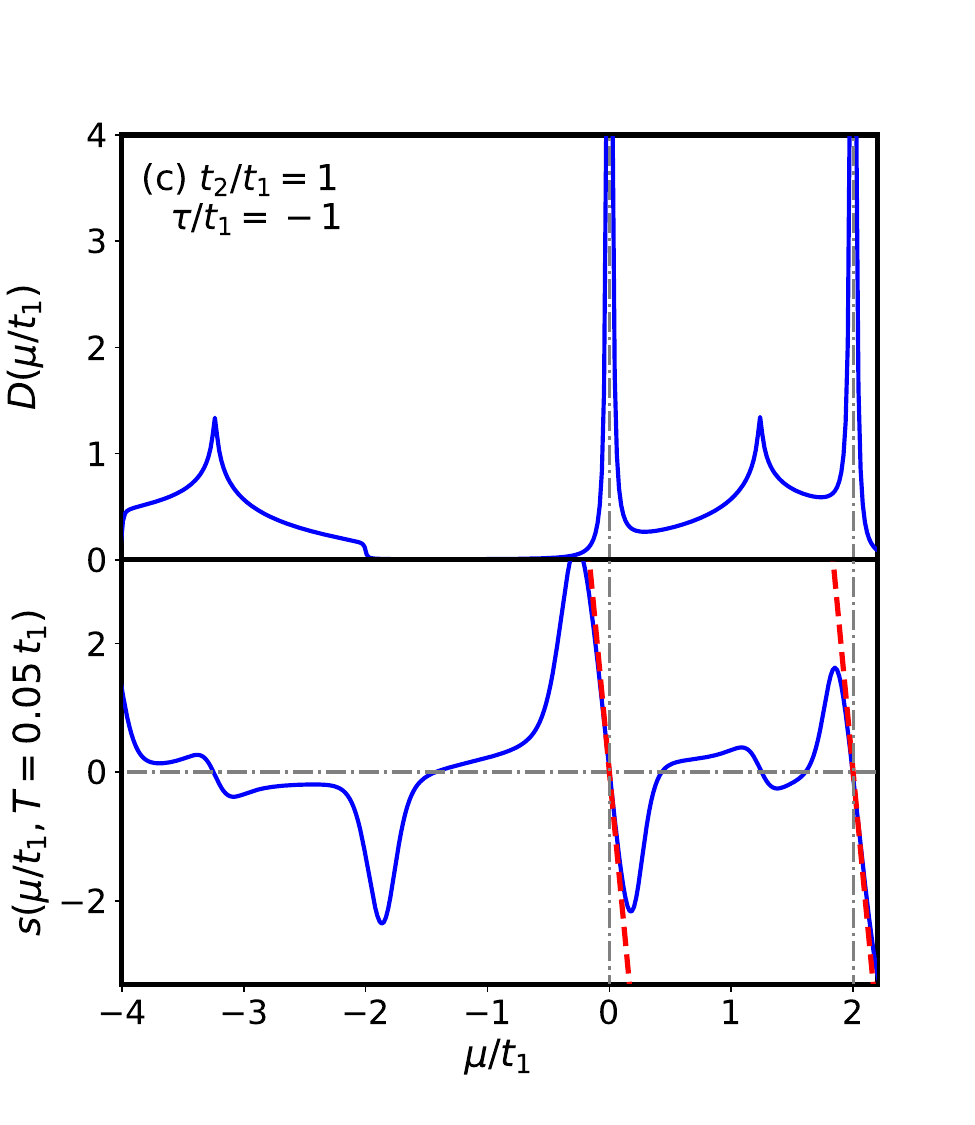}
	\caption{Upper plot in each panel shows the numerically calculated DoS with level broadening parameter $\Gamma=0.01 t_1$ as a function of chemical potential. Lower plots show the entropy per particle compared with the analytic expressions for logarithmic (green dash-dotted), high-order vHs and flat band (red dashed) cases. The temperature is taken to be $T=0.05 t_1$, and the ratios of hopping parameters $t_2/t_1$ and $\tau/t_1$ are shown in each panel. The gray vertical lines show the exact positions of high-order vHs and flat bands in spectrum. The chemical potential is measured in units of $t_1$ hopping parameter. }
	\label{fig:Tgraphene-dos-entropy}
\end{figure}

The characteristic singularity in DoS for high-order vHs, that comes from flat lines in spectrum, is the same for all values of $t_2/t_1$ (we firstly analyze the case of $\tau=0$), and is given by
\begin{align}
	D(\epsilon\approx t_1)=D_0 \begin{cases}
		|\epsilon-t_1|^{-1/2},& \epsilon<t_1 \\
		0,& \epsilon>t_1
	\end{cases},\quad D(\epsilon\approx -t_1)=D_0 \begin{cases}
	|\epsilon+t_1|^{-1/2},& \epsilon>-t_1 \\
	0, & \epsilon<-t_1
\end{cases}.
\end{align} 
This represent the limiting case of vHs asymmetry, since one of the coefficients $D_{+}$ and $D_{-}$ is zero. Also, this model has one of the strongest divergencies comparing to lattices classified in Ref.\cite{Chandrasekaran2020PRB}. In panels (a) and (b) of Fig.\ref{fig:Tgraphene-dos-entropy} we plot numerically calculated density of states and the corresponding differential entropy, together with analytic expressions \eqref{eq:entropy-power-alpha} (shown as red dashed lines). The temperature was chosen as  $T=0.05 \,t_1$ and the level broadening was set to be $\Gamma=0.01\,t_1$. We specially selected larger temperature that in Sec.\ref{sec:graphene} to check the applicability of analytic expressions when vHs are more pronounced. Notably, the results in Fig.\ref{fig:Tgraphene-dos-entropy} show very good agreement between Eq.\eqref{eq:entropy-power-alpha} and numerical results. As was discussed in Refs.\cite{Yamashita2013PRB,Oriekhov2021}, in the $t_2/t_1=3/2$ case the flat lines are separated from upper (lower) bands by a large gap, and in the case of $t_2/t_1=1$ there are three-band-touching points with two Dirac cones exactly at vHs level. Still, the analytic expression works very well because the integrals in numerator and denominator in the differential entropy definition \eqref{eq:entropy-definition-integral} are strongly dominated by vHs contributions. The plot range of analytic expression covers more than $[-4T,+4T]$ interval around vHs. Also, one should note that zeros of the differential entropy appear near DoS extrema, as was qualitatively discussed in Ref.\cite{Galperin2018}.

Also, the model contains logarithmic van Hove singularities that are related to usual saddle points in band spectrum. They are always placed near M points (middle points on the sides of BZ), and for $\tau=0$ case their energies are given by $\mp t_{1} \sqrt{1+4 (t_2/t_1)^{2}}$. In this case we plot the linearized expression of $s_{log}$, which allows one to identify the characteristic parameter $\epsilon_0$. In panels (a) and (b) we shown the comparison of linear curves with properly-selected $\epsilon_0$ parameter vs numerical results for the differential entropy.  We find relatively large value of parameter $\epsilon_0=9 t_1$. However, here the $s_{log}(\mu,T)$ does not fully describe the differential entropy because the curve does not pass through zero exactly at vHs point. One can relate this to the presence of additional large constant density of states that surround vHs as it is done in the end of Appendix \ref{appendix:integrals}. In fact, taking into account additional constant density of states, one finds the numerically observed shift of differential entropy curve from zero for logarithmic vHs, but not for symmetric high-order vHs. In this sense the symmetric high-order vHs and flat bands as limiting case demonstrate more stable qualitative features in $s(\mu,T)$ than other types of DoS singularities. 

Finally, we analyze the regime of so-called ``critical" value of next-nearest neighbor hopping parameter $\tau=-t_1$ (with $t_1=t_2$), for which the two flat bands in spectrum appear \cite{Nunes2020}. Such spectrum is shown in panel (d) of Fig.\ref{fig:Tgraphene-lattice}, and the corresponding DoS / differential entropy comparison is presented in Fig.\ref{fig:Tgraphene-dos-entropy}(c). The exact energies of flat bands are $\epsilon_{flat}=0,\,2t_1$. We plot the analytic expressions from Eq.\eqref{eq:entropy-flat-band} for both cases and find very good agreement with numerical results. Also we should note that the peaks in $s(\mu,T)$, that surround the flat band zero are the most pronounced features of the differential entropy in such regime. This is related to the fact that flat bands are placed near spectral gap and band edge, respectively. 

%\subsection{Signatures of changing dispersions in tight-binding model}
%\label{sec:changing-dispersion}

\section{Conclusions}
\label{sec:conclusions}

In the present paper we derived the analytic expressions for the differential entropy per particle near the usual logarithmic and high-order (power law) van Hove singularities as well as flat bands. 
The important characteristic of the obtained results for high-order vHs and flat bands is the universality of the differential entropy expressions - they do not contain material-specific constants apart from those characterizing the vHs type and the position of flat band. The obtained expression for the differential entropy near usual logarithmic vHs is not universal and contains a parameter that should be estimated from the effective theory which describes saddle point in particular material or from numerical data. 

For the case of high-order van Hove singularities the slope of the entropy curve and its shift from zero allow one to identify the type of vHs. In fact, one may expect that this approach of identifying vHs type can be more precise than measurement of DoS in the case of symmetric van Hove singularities. The reason is that the slope of the entropy curve as a function of $\mu/T$ immediately gives the divergence exponent. In addition, the obtained expressions show that the entropy near flat band has the largest slope compared to all physically possible high-order van Hove singularities. The slope for logarithmic singularities is expected to be even smaller. 

In Refs.\cite{Tsaran2017Nature,Galperin2018} it was noted that zeros of the entropy per particle correspond to the extrema of the DoS curve as a function of $\mu$. In addition, the logarithmic van Hove singularities manifest themselves as strong positive peak and negative dip structures. Our results fully support these conclusions and generalize them to the case of high-order van Hove singularities. In fact, for the symmetric high-order vHs and flat band the structure of the differential entropy is qualitatively the same, but the peaks and dips have much larger amplitude than that around logarithmic vHs level. For asymmetric high-order vHs the pronounced peak or dip may be absent depending on asymmetry type.

We performed numerical calculations of DoS and the differential entropy for several tight-binding models: graphene with fine-tuned next- and next-next-nearest neighbor hopping parameters, Lieb lattice and square-octagon lattice. Our analytic predictions show good agreement with numerical results. However, one important limitation of the obtained formulas is that the corresponding spectral feature in DoS should be well separated in energy from other possible features - other van Hove singularities, flat bands or large constant DoS. The region where the analytic formulas are expected to work well is $\pm T$ from vHs or flat band energy level. In the case when other spectral features are presented nearby, one should derive combined expressions, taking into account the corresponding DoS of all of them. However, such results quickly become very complicated and go beyond the scope of our paper. One of interesting physical question that have to be addressed elsewhere is how the merging of several van Hove singularities with changing model parameters is manifested in the differential entropy.

\begin{acknowledgements}
	We are grateful to V. P. Gusynin, E. V. Gorbar, and S. G. Sharapov for fruitful discussions and critical reading of the manuscript. Yelizaveta Kulynych acknowledges a support by the National
	Research Foundation of Ukraine grant  (2020.02/0051)
	''Topological phases of matter and excitations
	in Dirac materials, Josephson junctions and magnets". D. O. Oriekhov acknowledges the support from the Netherlands Organization for Scientific Research
	(NWO/OCW) and from the European Research Council (ERC) under the European Union's Horizon 2020 research and innovation programme.
\end{acknowledgements}

\appendix
\section{Evaluation of entropy integrals through polylogarithm functions}
\label{appendix:integrals}
 We start with the integral for particle concentration for density of states describing the high-order van Hove singularity
\begin{align}\label{eq:n-alpha-mu-T}
	n_{\alpha}(\mu, T)=\int_{-\infty}^{\infty} \frac{D_{\alpha}(\varepsilon)}{\exp \left(\frac{\varepsilon-\mu}{T}\right)+1} d \varepsilon = C_2 T^{1-\alpha} \epsilon_0^{\alpha}\int_{-\infty}^{\infty} \bigg[D_{+}\Theta(\tilde{\epsilon})\left(\frac{1}{\tilde{\epsilon}}\right)^{\alpha}+D_{-}\Theta(-\tilde{\epsilon})\left(\frac{1}{|\tilde{\epsilon}|}\right)^{\alpha}\bigg]\frac{1}{\exp \left(\tilde{\epsilon}-\tilde{\mu}\right)+1} d \tilde{\epsilon}.
\end{align}
Here we introduced dimensionless variables $\tilde{\epsilon}=\epsilon/T$ and $\tilde{\mu}=\mu/T$ to simplify the notation.
%Again the cut-off parameter $\Lambda$ is introduced to describe the applicability for the effective DoS formula, but now for Eq.\eqref{eq:dos-power}.
Separating positive and negative regions of integration, we write:
\begin{align}
	Q_{\alpha}(\mu,\Lambda)&=\int_{-\infty}^{\infty} \bigg[D_{+}\Theta(\tilde{\epsilon})\left(\frac{1}{\tilde{\epsilon}}\right)^{\alpha}+D_{-}\Theta(-\tilde{\epsilon})\left(\frac{1}{|\tilde{\epsilon}|}\right)^{\alpha}\bigg]\frac{1}{\exp \left(\tilde{\epsilon}-\tilde{\mu}\right)+1} d \tilde{\epsilon}=\nn
	&=D_{+}\int_{0}^{\infty} \frac{1}{\tilde{\epsilon}^{\alpha}}\frac{1}{z\exp \left(\tilde{\epsilon}\right)+1} d \tilde{\epsilon}+D_{-}\int_{0}^{\infty} \frac{1}{\tilde{\epsilon}^{\alpha}}\frac{1}{z\exp \left(-\tilde{\epsilon}\right)+1} d \tilde{\epsilon},\quad z=e^{-\tilde{\mu}}\equiv e^{-\mu/T}.
\end{align}
In the second term from square brackets we made the change of variables $\tilde{\epsilon}\to-\tilde{\epsilon}$. After such replacement it becomes evident that the second integral with $D_{-}$ pre-factor requires proper regularization, since it diverges at the upper limit $\tilde{\epsilon}\to \infty$. This is because the denominator tends to $1$ at large $\tilde{\epsilon}$. To perform the calculation in proper way we note that the formula \eqref{eq:dos-power} has a finite range of applicability in energy domain around saddle point level. This in turn introduces a natural cut-off parameter for energy $E_0$, which after rescaling by temperature is used below as $\Lambda=E_0/T$. After the extraction of the singular part, the integrals become (we omit tilde in integration variable for simplicity):
\begin{align}\label{eq:Q-ans-alpha}
	Q(\mu,\alpha)&=D_{+}\frac{1}{z}\int_{0}^{\infty} \frac{1}{\epsilon^{\alpha}}\frac{1}{\exp \left(\epsilon\right)+(1/z)} d \epsilon+D_{-}\int_{0}^{\infty} \frac{1}{\epsilon^{\alpha}}\frac{\exp \left(\epsilon\right)}{z+\exp \left(\epsilon\right)} d \epsilon=\nn
	&=\frac{D_{+}}{z}\int_{0}^{\infty} \frac{1}{\epsilon^{\alpha}}\frac{1}{\exp \left(\epsilon\right)+(1/z)} d \epsilon+D_{-}\int_{0}^{\Lambda} \frac{1}{\epsilon^{\alpha}} d \epsilon-zD_{-}\int_{0}^{\infty} \frac{1}{\epsilon^{\alpha}}\frac{1}{z+\exp \left(\epsilon\right)} d \epsilon=
	\nn
	%&=\frac{1}{z}\frac{\mathrm{Li}_{1-\alpha}\left(-\frac{1}{z}\right)\Gamma(1-\alpha)}{-\frac{1}{z}}+\frac{\Lambda ^{1-\alpha}}{1-\alpha}-z\frac{\mathrm{Li}_{1-\alpha}\left(-z\right)\Gamma(1-\alpha)}{-z}=\nn
	&=-D_{+}\mathrm{Li}_{1-\alpha}\left(-\frac{1}{z}\right)\Gamma(1-\alpha)+\frac{D_{-}\Lambda ^{1-\alpha}}{1-\alpha}+D_{-}\mathrm{Li}_{1-\alpha}\left(-z\right)\Gamma(1-\alpha)
\end{align}
In the last line we used the Appell's integral for the polylogarithm function \cite{Whittaker2009},
\begin{align}
	\mathrm{Li}_{s}(z)=\frac{z}{\Gamma(s)} \int_{0}^{\infty} \frac{t^{s-1}\,\, d t }{e^{t}-z},\quad \Re s>0.
\end{align}
The condition for real part $\Re s>0$ is satisfied by the physical limit $0<\alpha<1$.
Substituting Eq.\eqref{eq:Q-ans-alpha} back into the expression for concentration of electrons, we find:
\begin{align}\label{eq:n-alpha-reg}
	n_{\alpha}(\mu,T)=C_2 T^{1-\alpha} \varepsilon_{0}^{\alpha}\Gamma(1-\alpha)\left[D_{-}\mathrm{Li}_{1-\alpha}\left(-z\right)-D_{+}\mathrm{Li}_{1-\alpha}\left(-\frac{1}{z}\right)\right]+C_2  \varepsilon_{0}^{\alpha}\frac{D_{-}E_0 ^{1-\alpha}}{1-\alpha},
\end{align}
where we specially separated the regularization constant out from brackets. This constant is independent of $\mu$ and $T$ and only depends on the parameters of particular system and $\alpha$. Thus, we rewrite this equation in a more formal way as
\begin{align}\label{eq:vHs-particle-concentration}
	&n_{\alpha}(\mu,T)=n_0(\alpha)+C_2 T^{1-\alpha} \varepsilon_{0}^{\alpha}\Gamma(1-\alpha)\left[D_{-}\mathrm{Li}_{1-\alpha}\left(-z\right)-D_{+}\mathrm{Li}_{1-\alpha}\left(-\frac{1}{z}\right)\right].
\end{align}
 Here we introduced $n_0$ as material-specific regularizing constant, which accounts for the filled levels deep under the Fermi level, and does not depend on $\mu$ and $T$. This constant is automatically canceled in derivatives presented in Eq.\eqref{eq:s-Maxwell-1}, so it can be safely ignored here.

Now the entropy per particle is calculated through the derivatives of concentration via Eq.\eqref{eq:s-Maxwell-1}. The constant $n_0=C_2  \varepsilon_{0}^{\alpha}\frac{D_{-}E_0 ^{1-\alpha}}{1-\alpha}$ is automatically canceled by derivatives. By noting that the dependence on $\mu$ enters $n_{\alpha}(\mu,T)$ only through variable $z=\exp(-\mu/T)$, we simplify the expression for entropy as follows:
\begin{align}\label{eq:A7-appendix}
	s&=\left(\frac{\partial n_{\alpha}}{\partial T}\right)_{\mu}\left(\frac{\partial n_{\alpha}}{\partial \mu}\right)_{T}^{-1}=\left[(1-\alpha)\frac{n_{\alpha}(\mu,T)-n_0}{T}+T^{1-\alpha}\p_{z}\left(\frac{n_{\alpha}(\mu,T)-n_0}{T^{1-\alpha}}\right)\p_{T}z\right]\left[T^{1-\alpha}\p_{z}\left(\frac{n_{\alpha}(\mu,T)-n_0}{T^{1-\alpha}}\right)\p_{\mu}z\right]^{-1}\nn
	&=\frac{(1-\alpha)\frac{n_{\alpha}(\mu,T)-n_0}{T}}{T^{1-\alpha}\p_{z}\left(\frac{n_{\alpha}(\mu,T)-n_0}{T^{1-\alpha}}\right)\p_{\mu}z}+\frac{\p_T z}{\p_{\mu}z}%=\frac{(1-\alpha)\frac{n(\mu,T)}{T}}{T^{1-\alpha}\p_{z}\left(\frac{n(\mu,T)}{T^{1-\alpha}}\right)\p_{\mu}z}
	.
\end{align}
 The last ratio of the derivatives in this expression gives the contribution $-\mu/T$ to the differential entropy. The derivative over $z$ has the form:
\begin{align}
	\p_{z}\left(\frac{n_{\alpha}(\mu,T)-n_0}{T^{1-\alpha}}\right)&=C_2\epsilon_0^{\alpha}\Gamma(1-\alpha)\p_z \left[D_{-}\mathrm{Li}_{1-\alpha}\left(-z\right)-D_{+}\mathrm{Li}_{1-\alpha}\left(-\frac{1}{z}\right)\right]=\nn
	&=C_2\epsilon_0^{\alpha}\frac{\Gamma(1-\alpha)}{z}\left[D_{-}\mathrm{Li}_{-\alpha}\left(-z\right)+D_{+}\mathrm{Li}_{-\alpha}\left(-\frac{1}{z}\right)\right].
\end{align}
Substituting this back into expression for entropy and canceling the equal factors, we find
\begin{align}
	s&=-\frac{(1-\alpha)T^{1-\alpha}\Gamma(1-\alpha)\left(D_{-}\mathrm{Li}_{1-\alpha}\left(-z\right)-D_{+}\mathrm{Li}_{1-\alpha}\left(-\frac{1}{z}\right)\right)}{T^{2-\alpha}\frac{\Gamma(1-\alpha)}{z}\left[D_{-}\mathrm{Li}_{-\alpha}\left(-z\right)+D_{+}\mathrm{Li}_{-\alpha}\left(-\frac{1}{z}\right)\right] \frac{z}{T}}-\frac{\mu}{T}\nn
	&=-(1-\alpha)\frac{D_{-}\mathrm{Li}_{1-\alpha}\left(-z\right)-D_{+}\mathrm{Li}_{1-\alpha}\left(-\frac{1}{z}\right)}{D_{-}\mathrm{Li}_{-\alpha}\left(-z\right)+D_{+}\mathrm{Li}_{-\alpha}\left(-\frac{1}{z}\right) }-\frac{\mu}{T}.
\end{align}
This expression appears in the main text, see Eq.\eqref{eq:entropy-power-alpha}. 

Next we perform the calculation of entropy for logarithmic density of states \eqref{eq:dos-log}, that describes ordinary van Hove singularity. Again we start with the particle concentration and note the useful relation with Eq.\eqref{eq:n-alpha-mu-T} for $n_{\alpha}(\mu,T)$:
\begin{align}
	n_{log}(\mu, T)=\int\limits_{-\infty}^{\infty} \frac{D_{log}(\varepsilon)d \varepsilon}{\exp \left(\frac{\varepsilon-\mu}{T}\right)+1} =C_1 \frac{\p}{\p\alpha}\left[\frac{n_{\alpha}(\mu,T,D_{+}=D_{-})}{C_2}\right]_{\alpha\to 0}.
	%T \int\limits_{-\infty}^{\infty} \log\frac{\tilde{\epsilon}_0}{|\tilde{\epsilon}|}\frac{1}{\exp \left(\tilde{\epsilon}-\tilde{\mu}\right)+1} d \tilde{\epsilon}.
\end{align}
Here we used the algebraic relation $\p_\alpha (\epsilon_0/|\epsilon|)^{\alpha}|_{\alpha\to 0}=\log(\epsilon_0/|\epsilon|)$. Using this relation and taking into account that $n_{\alpha}(\mu,T)$ under derivative should be properly regularized (see Eq.\eqref{eq:n-alpha-reg}), we can express the differential entropy as follows:
\begin{align}
	s_{\log}(\mu, T)=\left[\frac{\partial}{\partial \alpha}\left(\frac{\partial n_{\alpha}}{\partial T}\right)_{\mu}\right]_{\alpha\to 0}\left[\frac{\partial}{\partial \alpha}\left(\frac{\partial n}{\partial \mu}\right)_{T}\right]_{\alpha\to 0}^{-1}.
\end{align}
In this equation we changed the order of derivatives, which would allow us to cancel material-dependent regularization constants. Also it reduced the problem to evaluating the derivatives of previously obtained result for particle concentration in the case of high-order saddle point \eqref{eq:n-alpha-reg}. Evaluating the derivatives, we find:
\begin{align}\label{eq:log-s-parts}
	&\left(\frac{\partial n}{\partial T}\right)_{\mu}=-C_{1}\left(\frac{\mu}{T}\left(\mathrm{Li}^{(1,0)}\left(0,-e^{-\frac{\mu}{T}}\right)+\mathrm{Li}^{(1,0)}\left(0,-e^{\frac{\mu}{T}}\right)\right)+\mathrm{Li}^{(1,0)}\left(1,-e^{-\frac{\mu}{T}}\right)-\mathrm{Li}^{(1,0)}\left(1,-e^{\frac{\mu}{T}}\right)+\right.\nn
	&+\left.\left(\log \left(\frac{\varepsilon_{0}}{T}\right)+\gamma-1\right)\left(\log \left(e^{-\frac{\mu}{T}}+1\right)-\log \left(e^{\frac{\mu}{T}}+1\right)\right)+\frac{\mu}{T}\left(\log \left(\frac{\varepsilon_{0}}{T}\right)+\gamma\right)\right),\\
	&\left(\frac{\partial n}{\partial \mu}\right)_{T}=C_{1}\left(\mathrm{Li}^{(1,0)}\left(0,-e^{-\frac{\mu}{T}}\right)+\mathrm{Li}^{(1,0)}\left(0,-e^{\frac{\mu}{T}}\right)+\gamma- \log \left(\frac{\varepsilon_{0}}{T}\right)\right).
\end{align}
 Here $\gamma\approx 0.5772$ is the Euler constant. Combining these two expressions, we arrive at Eq.\eqref{eq:entropy-log} in the main text. 

To explore the effects of other DoS features near vHs level we preform the calculation for density of states which includes constant term $D_0$ additionally to vHs contributions from Eqs.\eqref{eq:dos-log} and \eqref{eq:dos-power}. By noting that the constant DoS corresponds to the limit $\alpha\to 0$ with $D_+=D_-=D_0$ of Eq.\eqref{eq:dos-power}, we can use the previous results of integration from this Appendix. For the high-order vHs case, taking $D(\epsilon)=D_{\alpha}(\epsilon)+C_2 D_0$ we find the concentration
\begin{align}
	n_{\alpha,\,const}(\mu,T)=n_{\alpha}(\mu, T)+ n_{\alpha\to 0}(\mu, T,D_{\pm}\to D_0).
\end{align}
Substituting this into Eq.\eqref{eq:s-Maxwell-1}, we find the modified expression differential entropy:
\begin{align}
	s&=\frac{\left(\frac{\partial n_{\alpha,\,const}}{\partial T}\right)_{\mu}}{\left(\frac{\partial n_{\alpha,\,const}}{\partial \mu}\right)_{T}}=\frac{(1-\alpha)\frac{n_{\alpha}(\mu,T)-n_0}{T}+T^{1-\alpha}\p_{z}\left(\frac{n_{\alpha}(\mu,T)-n_0}{T^{1-\alpha}}\right)\p_{T}z+\frac{n_{\alpha\to 0}(\mu,T)-n_0}{T}+T\p_{z}\left(\frac{n_{\alpha\to 0}(\mu,T)-n_0}{T^{1}}\right)\p_{T}z}{T^{1-\alpha}\p_{z}\left(\frac{n_{\alpha}(\mu,T)-n_0}{T^{1-\alpha}}\right)\p_{\mu}z+T\p_{z}\left(\frac{n_{\alpha\to 0}(\mu,T)-n_0}{T}\right)\p_{\mu}z}.
\end{align}
Substituting the results of differentiation and simplifying the expressions as in Eq.\eqref{eq:A7-appendix}, we find:
\begin{align}\label{eq:s-modified-A16}
	s&=-\frac{(1-\alpha)T^{1-\alpha} \varepsilon_{0}^{\alpha} \Gamma(1-\alpha)\left[D_{-} \operatorname{Li}_{1-\alpha}(-z)-D_{+} \operatorname{Li}_{1-\alpha}\left(-\frac{1}{z}\right)\right]+T \left[D_{0} \operatorname{Li}_{1}(-z)-D_{0} \operatorname{Li}_{1}\left(-\frac{1}{z}\right)\right]}{T^{1-\alpha}\varepsilon_{0}^{\alpha} \Gamma(1-\alpha)\left[D_{-} \mathrm{Li}_{-\alpha}(-z)+D_{+} \mathrm{Li}_{-\alpha}\left(-\frac{1}{z}\right)\right]+T\left[D_{0} \mathrm{Li}_{0}(-z)+D_{0} \mathrm{Li}_{0}\left(-\frac{1}{z}\right)\right]}-\frac{\mu}{T}. 
\end{align}
The main conclusion that can be drawn from this expression is that now behavior does not depend only on $\mu/T$ relation, but contains additional temperature dependence in first term. To perform further analysis, we reduce the complexity of expression by expanding it around $\mu=0$ for symmetric vHs case $D_{+}=D_{-}$:
\begin{align}
	s\approx -\frac{\alpha \mu }{T\left(1+\frac{D_{0} \epsilon_0^{-\alpha } T^{\alpha}}{2\alpha\left(2^{\alpha +1}-1\right) D_{+} \Gamma (-\alpha ) \zeta
			(-\alpha )}\right)}.
\end{align}
Notably, the differential entropy curve will still pass through zero at $\mu=0$, but now the slope will be modified and depend on temperature as well as other materials parameters such as $D_{+},\,D_{0}$ and $\epsilon_0$. If one takes sufficiently low temperatures, such that $\epsilon_0\gg T$, the correction to the slope will be small.

Performing the same calculation for logarithmic van Hove singularity, we find the following approximate expression at small around vHs level $\mu\ll T$ (technical details of calculations are the same as in Eqs.\eqref{eq:log-s-parts} and \eqref{eq:s-modified-A16}):
\begin{align}
	s=D_0 \left(\frac{1}{\log \left(\frac{2
			\varepsilon }{\pi  T}\right)+\gamma
	}+T\right)-\frac{  D_0+1}{ \log
		\left(\frac{\varepsilon_0}{  T}\right)+\gamma+\log
		\left(\frac{2}{\pi}\right)
		}\frac{\mu}{T}.
\end{align}
Now the first $\mu$-independent term appears, which describes the shift of curve from zero at $\mu$, as was observed numerically in Sec.\ref{sec:t-graphene}.

\section{Estimation of $\epsilon_0$ parameter for logarithmic vHs DoS in graphene with NNN hopping parameters}
\label{appendix:epsilon-0}
In this Appendix we show how the parameter $\epsilon_0$ that enters DoS for logarithmic vHs \eqref{eq:dos-log} can be estimated from the series expansion coefficients of dispersion relation near saddle point. For this purpose we use the example of graphene with NNN hopping parameters analyzed in Sec.\ref{sec:graphene}. In the lower band given by $\epsilon_{-}(\vec{k})$ in Eq.\eqref{eq:graphene-NNN-spectrum} one finds the logarithmic van Hove singularity at M points with energy $\varepsilon_{-}\left(\boldsymbol{k}=\boldsymbol{M}_{1}\right)=-\left|t_{1}-3 t_{3}\right|+2 t_{2}$. Expanding the dispersion up to fourth order in  around $\vec{M}_1$ point and taking tight-binding parameters used for calculations $t_2=0.2 t_1$, $t_3 = 0.15 t_1$, we find 
\begin{align}
	\epsilon_{-}(\vec{k}=\vec{M}_1+\vec{q})\approx -\frac{3}{20}t_1+\frac{3}{55} a^2 t_1 \left(11 q_x^2-96 q_y^2\right)+a^4 t_1\left(\frac{5751  q_x^2 q_y^2}{3872}-\frac{21}{320}  q_x^4+\frac{1463373 q_y^4}{85184}\right).
\end{align}
By noting that the fourth-order term is strongly dominated by the last fraction in brackets for nearly all $\vec{q}$, we can use the results of Ref.\cite{Yuan2019Nature} to estimate the $\epsilon_0$ parameter. For the dispersion of the form 
\begin{align}
	\epsilon=\epsilon_{vHs}+\alpha a^2 q_x^2 - \beta a^2 q_y ^2 + \kappa a^4 q_4^2
\end{align}
one finds the leading term in DoS $D(\epsilon)\sim \log \epsilon_0/|\epsilon-\epsilon_{vHs}|$ with $\epsilon_0 = (8 \beta^2 / \kappa) t_1$. Substituting our parameters we find $\epsilon_0 \approx 12.8 t_1$, which is close to one estimated numerically in Sec.\ref{sec:graphene}.

\bibliography{entropy_van_Hove_bib}

\end{document}